\documentclass[10pt,
a4paper,
twocolumn
]{article}
\pdfoutput=1
\RequirePackage[labelsep=space,tableposition=top,font=small]{caption}
\usepackage{subcaption}
\usepackage{booktabs} 
\usepackage{multirow}
\usepackage{siunitx} 
\usepackage{enumitem}
\usepackage{appendix}

\usepackage[left=2cm,right=2cm,top=2cm,bottom=2cm]{geometry}
\usepackage{amsmath}
\usepackage{amssymb}
\usepackage{braket}
\usepackage{mathtools}
\usepackage{graphicx}
\usepackage{verbatim}
\usepackage{textcomp}
\usepackage{url}
\usepackage{cprotect}
\usepackage{tikz}
\usepackage{color}
\usepackage{pgfgantt}
\usepackage{pdflscape}
\usepackage{slashed}
\usepackage[numbers]{natbib}
\usepackage{caption}
\usepackage{subcaption}
\usepackage{floatrow}
\usepackage{array}

\newcolumntype{.}{D{.}{.}{-1}}
\newcommand{\dd}{\mathrm{d}}
\newcommand{\ii}{\mathrm{i}}

\newcommand{\bfk}{\mathbf{k}}
\newcommand{\bfq}{\mathbf{q}}

\newcommand{\bfp}{\mathbf{p}}

\newcommand{\bfr}{\mathbf{r}}

\newcommand{\nup}{N_\uparrow}
\newcommand{\ndo}{N_\downarrow}

\newcommand{\expe}[0]{\mathrm{e}}

\newcommand{\EF}[0]{E_\mathrm{F}}

\newcommand{\kF}[0]{k_\mathrm{F}}

\newcommand{\abs}[1]{\left\lvert #1 \right\rvert}
\newcommand{\figref}[1]{Fig.~\ref{#1}}
\newcommand{\eqnref}[1]{Eq.~(\ref{#1})}

\newcommand{\bpsi}[0]{\bar{\psi}}
\newcommand{\Nq}[0]{N_q}

\definecolor{pu}{RGB}{200,50,200}
\definecolor{gr}{RGB}{0,150,0}
\definecolor{bl}{RGB}{68,34,200}
\definecolor{re}{RGB}{200,34,68}

\begin{document}
\title{Temporal fluctuation induced order in conventional superconductors}
\author{D.C.W.~Foo$^1$ \and G.J.~Conduit$^1$}
\date{$^1$\emph{Cavendish Laboratory, J.J. Thomson Avenue, Cambridge, CB3 0HE, 
United Kingdom}\\[2ex]
\today}

\twocolumn[
  \begin{@twocolumnfalse}
    \maketitle
    \begin{abstract}
Communal pairing in superconductors introduces variational freedom for Cooper pairs to share fermions. Temporal oscillations of the superconducting gap entropically drive communal pairing through the order by disorder phenomenology, stabilising a finite momentum space width of the superconducting gap that increases with interaction strength, creating a smooth evolution from the weakly interacting BCS state to the strongly interacting BEC state.
    \end{abstract}
  \end{@twocolumnfalse}
]
\setlength\arraycolsep{1pt}
\section{Introduction}\label{sec:intro}
The microscopic description of superconductivity by Bardeen, Cooper, and Schrieffer (BCS)~\cite{bcs} is one of the historic milestones of condensed matter physics, accurately describing a host of materials~\cite{supconelemrev,supconcomprev,consupconhight,consupconquacrys}, and serving as the foundation for numerous theoretical extensions and numerical studies, such as Eliashberg theory~\cite{eliashberg}, FFLO theory~\cite{ffloff,fflolo}, breached superconductivity~\cite{breachsupercon1,breachsupercon2,breachsupercon3}, the T-matrix formulation of the BEC-BCS crossover~\cite{2dfgtmat}, quantum Monte Carlo studies of the weakly interacting~\cite{qmcbcs} and unitarity limits~\cite{qmcbecbcs}, studies on the effects of mass imbalances~\cite{mimba1,mimba2}, 3-body effects~\cite{3bodbcsbec} and the more recent communal pairing theory~\cite{multipartsupconfew,commupair}. Central to the usual formulation of BCS theory is the assumption that the Cooper pairs condense only in the zero net momentum state, an assumption that is challenged by communal pairing theory~\cite{multipartsupconfew,commupair,commupairnum}.

Communal pairing theory as originally derived~\cite{multipartsupconfew} showed that it is energetically favourable for Cooper pairs to share fermions. By considering the quantities $N_\sigma$ with $\sigma\in\{\uparrow,\downarrow\}$, where $N_\uparrow$ is the number of up-spin fermions any particular down-spin fermion is paired with, and \emph{vice versa}, communal pairing theory predicts an optimal ratio of communal state indices of $\nup/\ndo=1$ for the spin-balanced BCS system and $\nup/\ndo\neq1$ in a spin-imbalanced system~\cite{commupair,commupairnum}. The central paradigm shift that a finite gap may be present at non-optimal pairing momenta allows $\nup$ and $\ndo$ to be greater than 1. It is therefore natural to ask whether as interactions get stronger in a spin-balanced BCS system, multiple Cooper pairs will share fermions, $\nup>1$ and $\ndo>1$, to increase correlations. The variational principle ensures that the inclusion of additional freedom to form correlations will certainly not increase the ground state energy so can only lower it.

This paper explores the extent of communality on spin-balanced systems. We do this by extending BCS theory, complementary to other additional effects, such as retardation as in Eliashberg theory~\cite{eliashberg} or induced Gor'kov-Melik-Barkhudarov interactions (GMB)~\cite{gmb,gmbdisentangle}. We will focus our discussion on 2D systems as communality is predicted to be enhanced in low dimensions~\cite{multipartsupconfew,commupair,commupairnum} and because the results may be derived analytically here, but will also derive equivalent 3D results where possible. We will also be pre-emptively setting $\nup=\ndo=\Nq$ to reflect the fact that the system is spin-balanced.

In the next section we briefly recap conventional superconductivity from a field theoretic perspective and note the main difficulty with an exact treatment. Section~\ref{sec:onecha} analyses single superconducting channels, making a distinction between static and oscillating channels before we combine these results into a minimally coupled model of multiple active superconducting channels in Section~\ref{sec:mulcha}, making clear the connection to the BEC-BCS crossover. Conclusions are summarised in Section~\ref{sec:conclusion}.

\section{Quantum action}
To start our analysis from a secure theoretical footing, we analyse the quantum partition function of a fermion gas with attractive contact interactions of strength $g$, $\mathcal{Z}=\int\mathcal{D}\bpsi\mathcal{D}\psi\exp(-S[\bpsi,\psi])$, where
\begin{align*}
S[\bpsi,\psi]=\int\dd\tau\dd x\bigg[\sum_{\sigma}\bpsi_\sigma(\partial_\tau\!-\!\frac{\nabla^2}{2m}-&\mu)\psi_\sigma-g\bpsi_\uparrow\bpsi_\downarrow\psi_\downarrow\psi_\uparrow\bigg],
\end{align*}
is the quantum action, $\psi$ is a Grassman field with $\bpsi$ its conjugate, $\tau$ the imaginary time goes from 0 to $\beta$ the inverse temperature, and $\sigma\in\{\uparrow,\downarrow\}$ denotes the spin-species. The fermions are of equal mass $m$ and we work in Hartree units so $\hbar=k_\mathrm{B}=1$. A Hubbard-Stratonovich decoupling in the Cooper channel yields the modified action
\begin{align*}
S[\bpsi,\psi,\Delta^*\!,\Delta]=\int\dd\tau&\dd x\bigg(\sum_{\sigma}\bpsi_\sigma(\partial_\tau\!-\!\frac{\nabla^2}{2m}-\mu)\psi_\sigma\\&-\Delta\bpsi_\uparrow\bpsi_\downarrow-\Delta^*\psi_\downarrow\psi_\uparrow+\frac{\Delta^*\Delta}{g}\bigg),
\end{align*}
where the gap parameter is defined as $\Delta\equiv g\langle\psi_\downarrow\psi_\uparrow\rangle$ and is a function of both space $x$ and time $\tau$. The Fourier transform of the gap is therefore generically a function of the pair momentum $\bfq$ and the frequency $\Omega$, which label the various superconducting channels. In the weakly interacting limit, the gap is known to be isotropic and static and therefore the Fourier transform is a delta function in the momentum-frequency domain. However, in the strongly interacting limit, approaching the BEC-BCS crossover, communal pairing allows Cooper pairs to share fermions~\cite{multipartsupconfew}. We note there is an analogy to Cooper pairs becoming confined in real space, which should correspond to a widening of the gap in momentum space. It is this width that is the central concern of this paper, and so we Fourier transform to momentum and frequency space to obtain the action
\begin{align*}
S[&\bpsi,\psi,\Delta^*\!,\Delta]=\beta\sum_{\bfk,\omega,\sigma}\bpsi_{\bfk,\omega,\sigma}(-\ii\omega+\xi_\bfk)\psi_{\bfk,\omega,\sigma}\\&-\beta\!\sum_{\bfk,\bfq,\omega,\Omega}\!\left(\!\Delta_{\bfq,\Omega}\bpsi_{\bfk+\tfrac{\bfq}{2},\omega+\tfrac{\Omega}{2},\uparrow}\bpsi_{-\bfk+\tfrac{\bfq}{2},-\omega+\tfrac{\Omega}{2},\downarrow}\!+\mathrm{h.c.}\!\right)\!\\&+\beta\sum_{\bfq,\Omega}\frac{\abs{\Delta_{\bfq,\Omega}}^2}{g},
\end{align*}
where $\bfk$ and $\bfq$ label momenta and pair momenta respectively, $\omega$ is a fermionic matsubara frequency, $\Omega$ is a bosonic matsubara frequency, $\xi_\bfk\equiv\lvert\bfk\rvert^2/2m-\mu$ is the free particle dispersion less the chemical potential, $\Delta_{\bfq,\Omega}=g\langle\sum_{\bfk,\omega}\psi_{\bfk+\tfrac{\bfq}{2},\omega+\tfrac{\Omega}{2},\downarrow}\psi_{-\bfk+\tfrac{\bfq}{2},-\omega+\tfrac{\Omega}{2},\uparrow}\rangle$ are the Fourier components of the gap function and h.c. denotes the Hermitian conjugate. 

We will ultimately consider communal pairing through multiple channels but to lay the foundation of the analysis, and connect to standard BCS theory, we will first decouple through a single $\Delta_{\bfq,\Omega}$ channel.

\section{Decoupling in a single channel}\label{sec:onecha}
We first follow the standard BCS prescription to permit each fermion to be paired with only one opposite spin fermion. Therefore, only one superconducting channel $\Delta_{\bfq,\Omega}$ is nonzero, revealing a key difference between the channels where $\Omega=0$ and $\Omega\neq0$, namely that while the action of the static channels is fully real, the oscillating channels have a complex action indicating a finite lifetime of the Cooper pairs. These different situations are dealt with in Subsections \ref{subsec:ssc} and \ref{subsec:finom} respectively. These decoupled expressions for the action will in Section~\ref{sec:mulcha} be combined to provide a full action where each fermion may be paired with every other.

With $\Delta=0$ except at a specific $\bfq$ and $\Omega$, the momentum sum in the three point interaction terms is simplified. The action is then
\begin{align*}
&S_\bfq[\bpsi,\psi,\Delta^*\!,\Delta]=\beta\frac{\abs{\Delta_{\bfq,\Omega}}^2}{g}\\&+\beta\sum_{\bfk,\omega}
\boldsymbol{\bpsi}_{\substack{\bfk,\bfq\\\omega,\Omega}}^{\mathrm{T}}
\begin{pmatrix}
G^{-1}_{\bfk+\tfrac{\bfq}{2},\omega+\tfrac{\Omega}{2},\uparrow} && -\Delta_{\bfq,\Omega}\\
-\Delta_{\bfq,\Omega}^* && G^{-1}_{-\bfk+\tfrac{\bfq}{2},-\omega+\tfrac{\Omega}{2},\downarrow}
\end{pmatrix}
\boldsymbol{\psi}_{\substack{\bfk,\bfq\\\omega,\Omega}}
,
\end{align*}
where $\boldsymbol{\bpsi}_{\substack{\bfk,\bfq\\\omega,\Omega}}^{\mathrm{T}}\equiv\begin{pmatrix}\bpsi_{\bfk+\tfrac{\bfq}{2},\omega+\tfrac{\Omega}{2},\uparrow}&&\psi_{-\bfk+\tfrac{\bfq}{2},-\omega+\tfrac{\Omega}{2},\downarrow}\end{pmatrix}$ and the inverse propagator $G^{-1}_{\bfp,\nu,\sigma}\equiv\sigma(-\ii\nu+\xi_\bfp)$. The fermion fields can then be integrated out to obtain the effective action
\begin{align}
&S_{\bfq,\Omega}[\Delta^*\!,\Delta]=\beta\frac{\abs{\Delta_{\bfq,\Omega}}^2}{g}\nonumber\\&-\sum_{\bfk,\omega}\ln\left(\!1\!-\!\abs{\Delta_{\bfq,\Omega}}^2 G_{\bfk+\tfrac{\bfq}{2},\omega+\tfrac{\Omega}{2},\uparrow}G_{-\bfk+\tfrac{\bfq}{2},-\omega+\tfrac{\Omega}{2},\downarrow}\right).
\label{action}
\end{align}
Far below the critical temperature, and for $q^2/2m+\Omega^2/4\mu<\abs{\Delta_{\bfq,\Omega}}^2/\mu$, we may perform the Matsubara summation to obtain
\begin{align}
S_{\bfq,\Omega}[\Delta^*\!,\Delta]=\beta&\frac{\abs{\Delta_{\bfq,\Omega}}^2}{g}-\beta\sum_{\bfk}\left(E_{\bfk,\bfq,\Omega}-\epsilon_{\bfk,\bfq,\Omega}\right)\nonumber\\&+\sum_{\bfk}\cosh\frac{\beta kq\cos\theta}{2m}\expe^{-\beta E_{\bfk,\bfq,\Omega}},
\label{divaction}
\end{align}
where $E_{\bfk,\bfq,\Omega}\equiv\sqrt{\abs{\Delta_{\bfq,\Omega}}^2+\epsilon_{\bfk,\bfq,\Omega}^2}$, $\epsilon_{\bfk,\bfq,\Omega}\equiv\frac{k^2}{2m}+\frac{q^2}{8m}-\mu-\ii\tfrac{\Omega}{2}$, and $\theta$ is the angle between $\bfk$ and $\bfq$. The limit on the magnitude of $\bfq$ is heuristically where the kinetic energy of the Cooper pair center of mass overcomes the superconducting condensation energy and therefore breaks the pair. Likewise, the limit on $\Omega$ sets a maximum allowed frequency of temporal oscillations of the gap. This therefore limits the stability of a finite $\Delta_{\bfq,\Omega}$ solution. The final term shows that the action has a leading order temperature dependence of the form $f(\beta,q,\Omega)\expe^{-\beta\abs{\Delta}}$ where $f$ is some bounded function. The term tends to zero as $T\to0$.

The first and second terms that remain at zero temperature require the contact interaction strength $g$ be regularized to eliminate the ultraviolet divergence. We replace $g$ with the s-wave scattering length $a_{\mathrm{s}}$ via the formal substitution~\cite{2dscaadhikari86,2dscaverhaar84,fermigasreview}
\begin{align}
\frac{1}{g}=
\begin{cases}
mL^2\left(\frac{1}{2\pi}\ln(\kappa a)+\frac{1}{L^2}\sum_{\bfk}\frac{1}{k^2-\kappa^2}\right),&D=2,\\
mL^3\left(-\frac{1}{4\pi a_{\mathrm{s}}}+\frac{1}{L^3}\sum_{\bfk}\frac{1}{k^2}\right),&D=3,
\end{cases}\label{regu}
\end{align}
where $L$ is the system length, $a=\tfrac{\expe^{\gamma}}{2}a_{\mathrm{s}}$ is the scattering length scaled for convenience with $\gamma$ the Euler–-Mascheroni constant, and $\kappa$ is an unimportant momentum scale that will vanish once the regularisation procedure is carried out in full. The scattering length $a$ (or $a_{\mathrm{s}}$) may be directly controlled experimentally~\cite{ucatomcontrol1,ucatomcontrol2,ucatomcontrol3}. This formal substitution works to regularise the integrals of \eqnref{divaction} at any value of $\Omega$.

The ultraviolet divergence of \eqnref{divaction} is thus exactly cancelled, allowing us to take the sum over all $\bfk$. Additionally, the first term on the right hand side of the regularization in~\eqnref{regu} allows us to predict that the familiar exponential suppression factor $\expe^{-2/g\nu_\mathrm{F}}$ seen in the solid-state BCS gap will be replaced with $1/\kF a$ in 2D and $\expe^{\pi/2\kF a_\mathrm{s}}$ in 3D. 

The action in 2D can be resolved analytically as
\begin{align}
&S_{\bfq,\Omega}=-\frac{\beta mL^2}{4\pi}\Bigg[\mu_{q,\Omega}\left(\sqrt{\abs{\Delta_{\bfq,\Omega}}^2+\mu_{q,\Omega}^2}-\mu_{q,\Omega}\right)\nonumber\\&-\abs{\Delta_{\bfq,\Omega}}^2\ln\frac{ma^2}{\sqrt{\expe}}\left(\sqrt{\abs{\Delta_{\bfq,\Omega}}^2+\mu_{q,\Omega}^2}-\mu_{q,\Omega}\right)\Bigg],
\label{actionqo}
\end{align}
where $\mu_{q,\Omega}\equiv\mu-\frac{q^2}{8m}+\ii\tfrac{\Omega}{2}$ and we have neglected the finite temperature correction term. In 3D, the action may be evaluated in terms of an elliptic integral
\begin{align*}
S_{\bfq,\Omega}^{\mathrm{(3D)}}=-\frac{\beta mL^3}{\pi^2}\Bigg[&\frac{\pi\abs{\Delta_{\bfq,\Omega}^\mathrm{(3D)}}^2}{4a_\mathrm{s}}\\&+\sqrt{2m\mu_{q,\Omega}^\mathrm{(3D)\,5}}I_1\left(\frac{\abs{\Delta_{\bfq,\Omega}^\mathrm{(3D)}}}{\mu_{q,\Omega}^\mathrm{(3D)}}\right)\Bigg],
\end{align*}
where $I_1(z)\equiv\int_{0}^{\infty}\dd x x^2(\sqrt{z^2+(x^2-1)^2}-(x^2-1)-z^2/2x^2)$ is a dimensionless function. 

Now that we have derived an expression for the action we are well positioned to consider separately two cases, firstly the special case of $\Omega=0$ before extending this to the finite $\Omega$ system.

\subsection{Static single channel}\label{subsec:ssc}
The static action can be found by setting $\Omega=0$ in \eqnref{actionqo} to obtain
\begin{align*}
S_{\bfq}=-\frac{\beta mL^2}{4\pi}&\Bigg[\mu_{q}\left(\sqrt{\abs{\Delta_{\bfq}}^2+\mu_{q}^2}-\mu_{q}\right)\\&-\abs{\Delta_{\bfq}}^2\ln\frac{ma^2}{\sqrt{\expe}}\left(\sqrt{\abs{\Delta_{\bfq}}^2+\mu_{q}^2}-\mu_{q}\right)\Bigg],
\end{align*}
where we drop the $\Omega$ subscript entirely as it is understood to be zero. The action is a real number, confirming that the condensed phase is stable in time. We will now obtain the grand potential through the standard formula $\Phi=-T\ln\mathcal{Z}$, the gap $\Delta_\bfq$, and the chemical potential that promote a platform for our future analysis and allow us to compare to standard results.

\emph{Grand potential\quad}The grand potential $\Phi_{\bfq}$ is obtained directly from the action
\begin{align}
\Phi_{\bfq}=-\frac{mL^2}{4\pi}&\Bigg[\mu_{q}\left(\sqrt{\abs{\Delta_{\bfq}}^2+\mu_{q}^2}-\mu_{q}\right)\nonumber\\&-\abs{\Delta_{\bfq}}^2\ln\frac{ma^2}{\sqrt{\expe}}\left(\sqrt{\abs{\Delta_{\bfq}}^2+\mu_{q}^2}-\mu_{q}\right)\Bigg].
\label{bigphi}
\end{align}
As expected, the grand potential tends to that of the normal state when $\abs{\Delta_\bfq}=0$. In 3D the form of the grand potential similarly mirrors the 3D action without a factor of $\beta$

A subtlety that bears mention is that in both 2D and 3D the expression above requires $q^2/2m<\abs{\Delta_\bfq}^2/\mu$, that is when the additional kinetic energy of a Cooper pair is less than the condensation energy. Above that limit, the additional kinetic energy is sufficient to break the Cooper pairs and so the grand potential evaluates to zero identically.

\emph{Superconducting gap\quad}The gap is determined by requiring that the grand potential be stationary with respect to the gap, $\frac{\partial\Phi}{\partial \Delta_\bfq^*}=0$, giving the gap
\begin{align}
\Delta_\bfq=
\begin{cases}
\frac{1}{ma^2}\sqrt{1+2ma^2\mu_q}&q<\frac{2}{a}\\
0&q\geq\frac{2}{a}.
\end{cases}
\label{gap}
\end{align}
The gap in 3D meanwhile is the solution of the implicit equation

\begin{align}
-\frac{1}{\kF a_\mathrm{s}}=\frac{2}{\pi}\sqrt{\frac{\mu_{q}^\mathrm{(3D)}}{\EF}}I_2\left(\frac{\abs{\Delta_{\bfq}^\mathrm{(3D)}}}{\mu_{q}^\mathrm{(3D)}}\right),
\label{3Dgap}
\end{align}
where $I_2(z)\equiv\int_{0}^{\infty}\dd x(x^2/\sqrt{z^2+(x^2-1)^2}-1)$.

\emph{Solving for $\mu$\quad}The chemical potential $\mu$ is found from the equation $N=-\frac{\partial \Phi}{\partial\mu}$. In 2D it is then 
\begin{align*}
\mu=\EF\left(1-\frac{1}{\kF^2a^2}+\frac{q^2}{4\kF^2}\right),
\end{align*}
and that depends on the net momentum of the condensed Cooper pairs, owing to their kinetic energy.

In 3D we recast the BCS number equation~\cite{fermigasreview} to obtain
\begin{align}
1=\frac{3}{2}\sqrt{\frac{\mu_{q}^\mathrm{(3D)}}{\EF}}^3I_3\left(\frac{\abs{\Delta_{\bfq}^\mathrm{(3D)}}}{\mu_{q}^\mathrm{(3D)}}\right),
\label{3Dmu}
\end{align}
where $I_3(z)\equiv\int_0^\infty\dd x x^2\left[1-(x^2-1)/\sqrt{z^2+(x^2-1)}\right]$. Since the pair of coupled Eqns. (\ref{3Dgap}) and (\ref{3Dmu}) only depend implicitly on $\bfq$ through $\Delta_{\bfq}^\mathrm{(3D)}$ and $\mu_{q}^\mathrm{(3D)}$, we conclude that $\mu_{q}^\mathrm{(3D)}=\mu_\mathrm{BCS}^\mathrm{(3D)}$ and therefore $\mu^\mathrm{(3D)}=\mu_\mathrm{BCS}^\mathrm{(3D)}+\frac{q^2}{8m}$. At $q=0$ the forms of Equations (\ref{3Dgap}) and (\ref{3Dmu}) are indeed equivalent to those of the regularized BCS equations~\cite{fermigasreview}.

\emph{Weak interactions\quad}We study the weakly interacting limit in 2D by setting $\kF a\gg1$. In this limit, $\mu_q\approx\EF$ as expected and the gap reduces to $\Delta_{\bfq}\approx2\EF/\kF a$ from which we may extract the 3D analogue $\Delta_{\bfq,\mathrm{(3D)}}\sim\EF\expe^{-\pi/2\kF \abs{a_\mathrm{s}}}$ by inspection of the regularisation procedure in~\eqnref{regu}, which agrees with standard BCS theory~\cite{fermigasreview}. Similarly, we can state that in the weakly interacting limit the validity requirement $\frac{q^2}{2m}<\frac{\abs{\Delta_\bfq}^2}{\mu}$ takes the form $q<\frac{8\kF}{\expe^2}\expe^{-\pi/2\kF \abs{a_\mathrm{s}}}$.

\subsection{Oscillating single channel}\label{subsec:finom}
When $\Omega$ is nonzero, the action $S_{\bfq,\Omega}$ is complex. Expanding the 2D expression of~\eqnref{actionqo} in small $\Omega$ about our static $\Omega=0$ solution, we obtain
\begin{align*}
S_{\bfq,\Omega}=S_{\bfq,0}&-\ii\frac{\beta mL^2}{4\pi}\left(\sqrt{\abs{\Delta_{\bfq,\Omega}}^2+\mu_{q,0}^2}-\mu_{q,0}\right)\Omega\\
&+\mathcal{O}\left(\Omega^2\right).
\end{align*}
As $S_{\bfq,0}$ is real, the imaginary part of $S_{\bfq,\Omega}$ is linear in $\Omega$. All terms of order $\Omega^2$ and higher are proportional to at least the second power of $\abs{\Delta_{\bfq,\Omega}}$. The imaginary part of the action corresponds to the spontaneous decay rate of Cooper pairs condensed in this superconducting channel, $\Gamma_{\bfq,\Omega}^{\mathrm{sd}}=\abs{\Im S_{\bfq,\Omega}}$. In 3D, the action may be likewise expanded in the low-$\Omega$ limit to obtain an imaginary part of the action linear in $\Omega$.

In this section we have decoupled in a single channel, performing a BCS-like analysis of the resulting simplified action for general pair momentum $\bfq$ and frequency $\Omega$. Two broad conclusions follow from this analysis. Firstly, when $\Omega=0$, multiple $\bfq$ channels are stable with identical nonzero gap magnitude and equal grand potential. This should be observable in weakly interacting superconductors and can be understood as the system admitting a persistent supercurrent, provided the pair moves slowly enough that dissipation through quasiparticle excitation is not energetically feasible. Secondly, for $\Omega\ne0$, the action develops an imaginary term, limiting the lifetime of oscillating modes. This may be thought of as an inductance that promotes stability of gap-dependent macroscopic observables, such as the supercurrent.

\section{Multiple channels}\label{sec:mulcha}
The results of the previous section indicate that in principle, multiple channels are stable at any particular scattering length, which naturally begs the question of whether multiple channels coexist in the ground state. Such communal superconductivity has previously been analyzed and explored numerically in spin-imbalanced systems~\cite{commupair,commupairnum} and so it is natural to now look at spin-balanced systems. We will introduce the variational freedom to explore these multiple active channels with the communal parameter $\Nq=\sum_{\bfq}1$, the number of $\bfq$ channels with nonzero gap. Clearly, $\Nq$ is at least equal to 1 (standard BCS superconductor) and is bounded from above by one of two physical arguments that in 2D take the form,
\begin{align}
\Nq<\min\left(1+\frac{2N}{\kF^2 a^2},\frac{N}{2}\right),\label{nqmax}
\end{align}
where the first limit (weakly interacting) corresponds to all superconducting channels for which $q<2/a$ active and is exactly equivalent to the physical limit $q^2/2m<\abs{\Delta_\bfq}^2/\mu$ mentioned previously, with the second limit (strongly interacting) that there are only $N/2$ choices of pairing partner for each fermion. The crossover is at $\kF a=2$. We note for completeness that in 3D the first limit has the form $\Nq=1+Nq_{\mathrm{max}}(a_\mathrm{s})^3/2\kF^3$ where $q_{\mathrm{max}}(a_\mathrm{s})=\sqrt{2m/\mu(a_\mathrm{s})}\abs{\Delta_{\bfq}(a_\mathrm{s})}$ and the second limit is unchanged.

In addition to this new communal variational freedom, we will also consider the effect on the grand potential of the short-lived finite $\Omega$ modes, where quantum fluctuations of the temporally oscillating modes can contribute to driving communal ordering of the superconducting gap. This contribution will then be added to a minimal model of multiple active static $\Omega=0$ modes. We therefore calculate the total quantum partition function as $\mathcal{Z}=\mathcal{Z}_0\mathcal{Z}_{\Omega\neq0}$, where $\mathcal{Z}_0$ gives the multi-channel saddle point approximation and $\mathcal{Z}_{\Omega\neq0}$ accounts for temporal fluctuations of the various modes. The partition function allows us to find the grand potential and differentiate to obtain the expected value of the number of fermions shared between Cooper pairs, $\Nq$, and then explore the evolution of $\Nq$ as we approach the BEC-BCS crossover. We tackle the static and fluctuating contributions in order.

\subsection{Static channels}\label{subsec:statcha}
We first focus on the static $\Omega=0$ channels. The static part of the partition function has the form
\begin{align}
\mathcal{Z}_0=\exp\left[-\beta\frac{1}{\Nq}\sum_\bfq \Phi_\bfq(\Delta_\bfq,\mu)\right],
\label{Z0om}
\end{align}
which accounts for the long-lived, $\Omega=0$ channels where a $\abs{\Delta_{\bfq,0}}\neq 0$ mean-field solution is possible. The channels are coupled as they draw from the same reservoir of fermions with common chemical potential $\mu$. The averaging over modes may be understood in the context of the quantum action as considering each fermion as being paired to multiple opposite-spin fermions probabilistically, and all channels are equally weighted since the grand potential of each channel in the single channel decoupling is identical. The mean-field grand potential is then
\begin{align}
\Phi_0&=\frac{1}{\Nq}\sum_\bfq \Phi_\bfq(\Delta_\bfq,\mu)\nonumber\\
&=-\frac{\EF L^2(1+2ma^2\mu)^2}{4\pi\kF^2a^4}+\frac{\EF(\Nq-1)^2}{4\Nq\kF^2 a^2},
\label{mulphi0}
\end{align}
that is, a sum over the grand potentials of single channel superconductors. The contribution from oscillating channels does not depend on the magnitudes of the static channels and so we may find the superconducting gap in the same way as for the single channel, by requiring that $\Phi_0$ be extremised. The result is the same as in \eqnref{gap} except that $\mu$ is constant, making the gap vary with $q$ in contrast to the single channel picture where the gap had the same magnitude for all $q<2/a$. With the form of the gap, we may then evaluate the sum over $\bfq$, allowing us to split the energy contributions into the BCS grand potential (the first term) and $\Nq$--dependent communal correction that arises from the changing magnitude of the gap. For the physically realizable values of $\Nq$, the static grand potential is minimized at $\Nq=1$, the standard BCS result. We therefore turn to address the contributions from the oscillating channels to determine whether they can drive communal pairing with $\Nq>1$.

\subsection{Finite $\Omega$ plasma}\label{subsec:finomplas}
Having determined the static channel contribution, we may now consider the effect of fluctuations in $\Omega\neq0$ channels. We consider transitions of a Cooper pair from a stable static channel to a spontaneously decaying oscillating channel to obtain the occupation probability of an oscillating channel as $\Phi_{\bfq}\left(\abs{\Delta_{\bfq,\Omega}}\right)/(\Phi_{\bfq}\left(\abs{\Delta_{\bfq,0}}\right)+\Gamma_{\bfq,\Omega}^{\mathrm{sd}})$. The modification to the mean-field partition function accounting for the short-lived excitations is then
\begin{align}
\mathcal{Z}_{\Omega\neq0}&=1+\sum_{\bfq,\Omega\neq 0}\frac{\Phi_{\bfq}\left(\abs{\Delta_{\bfq,\Omega}}\right)}{\Phi_{\bfq}\left(\abs{\Delta_{\bfq,0}}\right)+\Gamma_{\bfq,\Omega}^{\mathrm{sd}}}\nonumber\\&\mkern+33mu +\frac{1}{2!}\left(\sum_{\bfq,\Omega\neq 0}\frac{\Phi_{\bfq}\left(\abs{\Delta_{\bfq,\Omega}}\right)}{\Phi_{\bfq}\left(\abs{\Delta_{\bfq,0}}\right)+\Gamma_{\bfq,\Omega}^{\mathrm{sd}}}\right)^2\nonumber\\&\mkern+33mu + \cdots
\nonumber\\&=\exp\left[\sum_{\bfq,\Omega\neq 0}\frac{\Phi_{\bfq}\left(\abs{\Delta_{\bfq,\Omega}}\right)}{\Phi_{\bfq}\left(\abs{\Delta_{\bfq,0}}\right)+\Gamma_{\bfq,\Omega}^{\mathrm{sd}}}\right].
\label{Zfinom}
\end{align}
The 1 accounts for the situation where no such excitations are present, the second term for when a single channel is excited, the third for when two are simultaneously excited and so on. 

The form of the grand potential may be taken from \eqnref{bigphi} and the spontaneous decay rate from Subsection~\ref{subsec:finom}. In order to evaluate the sums, it is necessary to express the gap magnitudes explicitly in terms of $\bfq$ and $\Omega$. Once again, we extremise the grand potential with respect to the $\Delta_{\bfq,\Omega}$ to obtain $\abs{\Delta_{\bfq,\Omega}}^2=\abs{\Delta_\bfq}^2-\Omega/ma^2$, which also allows us to explicitly compute the upper limit on $\Omega$ as $ma^2\Omega_{M}=\sqrt{8(1+ma^2\mu)-a^2q^2(1+2ma^2\mu)}-2$, which is positive for $qa<2$, the region we are interested in. From these relations we see that strong interactions, low $a$, drive oscillations of the gap.

With all this in place, we may now perform the summations to obtain the contribution to the partition function as

\begin{align}
\mathcal{Z}_{\Omega\neq 0}=\exp\Bigg[&\frac{2L^2\beta}{3\pi^2ma^4}\left(\!F(1)\!-\!F\left(\!1\!-\!\frac{\pi\Nq a^2}{L^2}\right)\!\right)\!\Bigg],
\label{mulphio}
\end{align}
where 

\begin{align*}
F(x)=\frac{1\!+\!x\left(1\!+\!2ma^2\mu\right)\left(2\sqrt{1\!+\!x\left(1\!+\!2ma^2\mu\right)}\!-\!3\right)}{1\!+\!2ma^2\mu}
\end{align*}
is a dimensionless function of the dimensionless variable $x$ characterising the $\Nq$ dependence of this part of the partition function. The term in the exponential is positive for $1<\Nq<1+L^2/\pi a^2$, that is, for all accessible values of $\Nq$, corresponding to an increase in the number of accessible microstates and thus an entropically driven decrease in the grand potential. The presence of temporally oscillating modes thus contributes an entropic term to the grand potential, increasing the number of accessible microstates and thereby reducing the grand potential. Qualitatively similar behaviour may be obtained in 3D. 

\subsection{Optimizing $\Nq$}\label{subsec:extr}
With the grand potential in place, we are well-positioned to determine $\Nq$. Combining the static and oscillating contributions of \eqnref{mulphi0} and \eqnref{mulphio} gives $\Phi$ as a function of $\Nq$, which may then be minimized to obtain the optimal number of stabilised communal pairing channels as

\begin{figure}[t]
	\centering
	\includegraphics[width=0.9\linewidth]{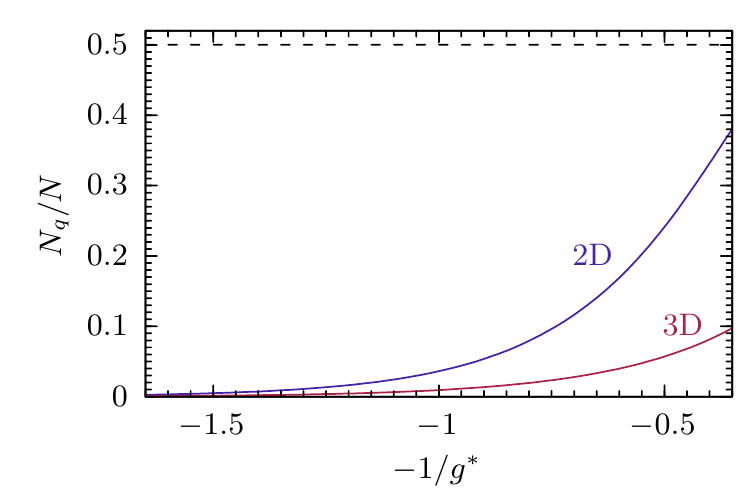}
	\caption{Plot of the ratio of $\Nq$ to $N$ as $N\to\infty$ as a function of dimensionless interaction strength $g^*$ in 2D (\textcolor{bl}{blue}) and 3D (\textcolor{re}{red}). In 2D, $-\frac{1}{g^*}=-\frac{1}{2}\ln(\kF a)$ while in 3D, $-\frac{1}{g^*}=\frac{3\pi}{8\kF a_\mathrm{s}}$. Interaction strength increases from left to right. The dotted black line indicates the theoretical maximum of $\frac{\Nq}{N}=\frac{1}{2}$.}
	\label{nqkfa}
\end{figure}

\begin{align}
\Nq=1+\frac{2N}{\kF^2 a^2}-\frac{\pi(32+\pi) N}{128(1+2ma^2\mu)\kF^2a^2},\label{nqrenorm}
\end{align}
that is, slightly fewer than the maximum permitted of \eqnref{nqmax}. Provided we remain in the weakly interacting regime $\kF a>1$ where our analysis is valid, the final correction term of \eqnref{nqrenorm} introduces a degree of negative feedback that ensures that $\Nq/N<1/2$ and we never encounter the hard physical limit. The BCS limit of $\Nq=1$ or, in the thermodynamic limit, $\Nq/N=0$, is recovered in the weakly interacting limit of $\kF a\to\infty$. We expect the above expression to be most correct in the regime where the scattering length $a$ is comparable to or less than the system size $L$ so that $\Nq>2$, with the BCS limit being a good description for even weaker interactions. The key role played by the temporal fluctuations here in determining the width in momentum space, and thus the real space structure, of the superconducting gap means that communal pairing in spin-balanced systems emerges as order by disorder.

The grand potential in 3D exhibits qualitatively similar behaviour, favouring $\Nq=1$ if not for the addition of the temporal fluctuation term, which instead promotes near maximal $\Nq$ provided interactions are weak, that is $\kF a_\mathrm{s}\to0^-$.

The emergence of communal pairing and increase of $\Nq$ is shown in \figref{nqkfa}, with $\Nq$ increasing smoothly as a function of the scattering length in both 2D and 3D. This is a marked difference from BCS theory which presupposes $\Nq=1$ at all interaction strengths. For ease of comparison, we have chosen to plot $\Nq$ as a function of the dimensionless interaction strength $g^*=\frac{Nv_0}{\EF}$ where $v_0$ is the inverse of the first term in the regularisation \eqnref{regu}, so $-\frac{1}{g^*}=-\frac{1}{2}\ln(\kF a)$ in 2D while in 3D, $-\frac{1}{g^*}=\frac{3\pi}{8\kF a_\mathrm{s}}$. The effect is stronger in 2D compared to 3D due to fluctuations being stronger in 2D.

\subsection{Adding gap fluctuations}

\begin{figure}[t]
	\centering
	\includegraphics[width=0.9\linewidth]{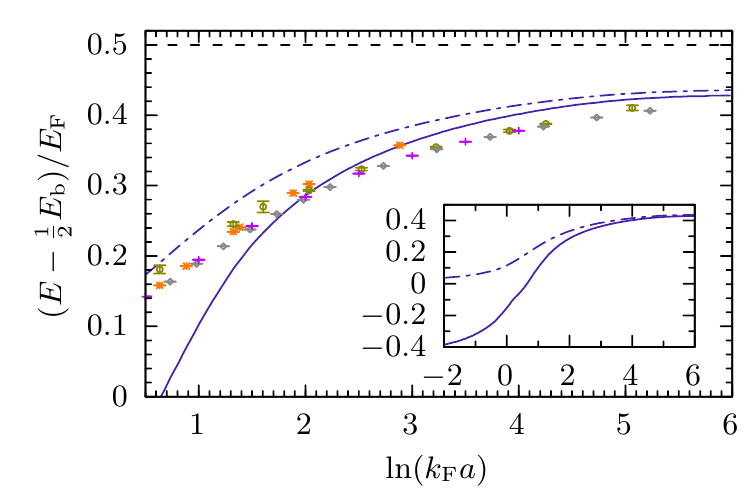}
	\caption{Plot of energy against $\ln(\kF a)$. The dashed line shows the BCS mean-field result, the dotted-dashed line shows the result obtained from treatment of Gaussian gap fluctuations using the T-matrix method and the solid line shows the T-matrix results with our communal correction. Various quantum Monte Carlo results are shown for comparison. The communal correction is seen to make up for a significant portion of the discrepancy between the T-matrix and Monte Carlo results at weak and intermediate interaction strengths. Inset: Bare (dotted-dashed) and communal corrected (solid) T-matrix solutions over a wider range of $\ln(\kF a)$.}
	\label{elnkfa}
\end{figure}

We have shown how temporal fluctuations and concomitant expansion of the phase space entropically stabilise a communal state. These considerations constitute a nontrivial extension of the original BCS theory that is nevertheless still a mean-field approach, and is therefore orthogonal to the usual treatment of Gaussian fluctuations of the order parameter, with which the T-matrix approach has had much success \cite{2dfgtmat}. It is therefore instructive to consider both sources of orthogonal fluctuation simultaneously by adding the difference between our obtained communal results and the traditional BCS mean field to the T-matrix results. 

This is shown in \figref{elnkfa} where we have plotted internal energy per particle against the interaction parameter $\ln(\kF a)$. The BCS mean field result is constant at 0.5, as seen by the dashed line, while the effects of Gaussian fluctuation of the order parameter obtained via the T-matrix approach are shown by the dotted-dashed line and are seen to consistently overestimate the energy calculated by quantum Monte Carlo methods \cite{2dunitarity1,2dunitarity2,dmcN1,2dunitarity4}. This disparity has previously been postulated as the GMB effect or beyond-quadratic fluctuations of the order parameter \cite{2dfgtmat}. Adding our correction to the T-matrix results gives the solid line which comes closer to the Monte Carlo results particularly around $\ln(\kF a)\approx2$. Furthermore, the quantum Monte Carlo results are more reliable in the intermediate interaction regime than the weakly interacting regime as the superconducting correlation length becomes smaller than the simulation cell length, making a correction in this intermediate regime particularly significant. We therefore contend that communal effects too may play an important role in the ground state. The inset of \figref{elnkfa} shows the same bare and communal corrected T-matrix results over the same range of $\ln(\kF a)$ as in~\cite{2dfgtmat}, where we see that at strong interactions our results greatly deviate from established results. This overshoot at $\ln(\kF a)<2$ is due to the aforementioned breakdown of assumptions at high interaction strength, and the system is now more correctly described as a weakly interacting 2D Bose gas~\cite{2dfgtmat}.


\subsection{Connection to BEC-BCS crossover}\label{subsec:bcsbeccon}

\begin{figure}[t]
	\centering
	\includegraphics[width=0.9\linewidth]{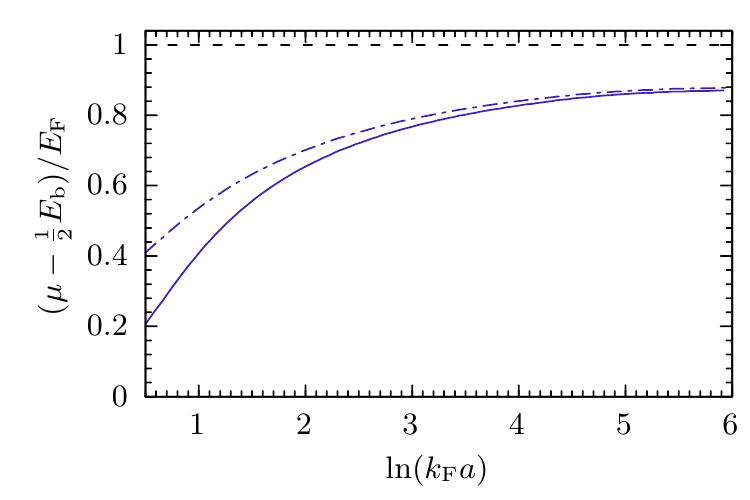}
	\caption{Plot of chemical potential against $\ln(\kF a)$. The dashed line shows the BCS mean-field result, the dotted-dashed line shows the result obtained from treatment of Gaussian gap fluctuations using the T-matrix method and the solid line shows the T-matrix results with our communal correction.}
	\label{mu}
\end{figure}

The increase in extent of communal pairing $\Nq$ and concomitant width of the gap in momentum space with increasing interaction strength points to a connection between communal superconductivity and the BEC state. The communal pairing state comprises many tightly bound, spatially localised Cooper pairs whose corresponding gap parameter is spread out in momentum space, analogous to the BEC state that comprises many tightly bound pairs of fermions. To probe this connection, we look to the chemical potential. Following the prescription of Subsection \ref{subsec:ssc}, we solve for $\mu$ and obtain

\begin{align*}
\frac{\mu}{\EF}=1-\frac{1}{\kF^2a^2}-\frac{8}{3\pi\kF^3a^3}+\mathcal{O}\left(\frac{1}{\kF^5a^5}\right).
\end{align*}
The first two terms are the BCS solution so the communal pairing correction is readily isolated as a reduction of the chemical potential, as seen in \figref{mu}. Starting from the non-interacting limit where $\mu=\EF$ as predicted by both BCS and communal pairing theory, as interactions get stronger $\mu$ decreases more quickly in communal pairing theory than in traditional pairing theory. The trends established in the communal state points towards a smooth evolution into the BEC regime, with a smooth confinement of more Cooper pairs with tighter spatial extent. The reduction of chemical potential persists even when incorporating the T-matrix analysis.

\begin{figure}[t]
	\centering
	\includegraphics[width=0.9\linewidth]{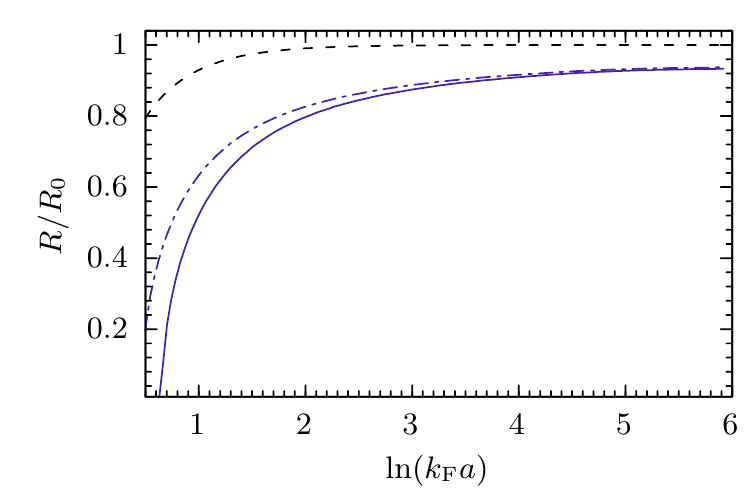}
	\caption{Plot of radius of a trapped interacting Fermi gas $R$ relative to the radius of a trapped noninteracting Fermi gas $R_0$ against $\ln(\kF a)$. The dashed line shows the BCS mean-field result, the dotted-dashed line shows the result obtained from treatment of Gaussian gap fluctuations using the T-matrix method and the solid line shows the T-matrix results with our communal correction.}
	\label{ra}
\end{figure}

This variation of chemical potential with interaction strength may be verified directly by experiment, for example by considering the radius of a trapped ultracold atomic gas. In the local density approximation, the chemical potential $\mu$ and density $n$ are related by $\mu\propto n^\gamma$ for some positive $\gamma$~\cite{fermigasreview} and so the radius of the trapped gas $R$ is where the local chemical potential vanishes, $\mu(R)\equiv\mu-V(R)=0$ where $V(\bfr)$ is the trapping potential. The radius of the trapped gas is thus a direct measure of the chemical potential by the relation $R\propto\sqrt{\mu}$. The full variation is shown in \figref{ra} where we see the change in radius is significant and should be readily observable in a cold atom gas. While the figure shows that our results indicate a collapse of the gas to a point at $\ln(\kF a)\approx0$, this occurs beyond the region of validity of our theory and is not expected to be experimentally observed.

\section{Discussion and conclusions}\label{sec:conclusion}
We have demonstrated the importance of communal corrections to the BCS theory by increasing variational freedom to include multiple superconducting modes. Partial occupancy of the temporally oscillating modes drive communal ordering of the superconducting gap, with each nonzero gap mode corresponding to a Cooper pair of net momentum $\bfq$ variationally lowering the grand potential, resulting in a favouring of multiple nonzero gap modes. Widening of the gap in momentum space, and the concomitant confinement of the Cooper pairs in real space with increasing interaction strength, points to a connection between communal superconductivity and the BEC-BCS crossover. Fluctuations of the gap itself were incorporated through the T-matrix analysis~\cite{2dfgtmat} resulting in a favourable comparison of the system energy with quantum Monte Carlo results.

The analysis focuses on how the partial occupancy of the temporal oscillating superconducting gap modes drive the emergence of communal order. This partial occupancy is driven by quantum fluctuations, that is by the uncertainty principle rather than temperature and so persist down to zero temperature where they affect the structure of the gap. We neglected the effect of density fluctuations that result in the GMB correction~\cite{gmb,gmbdisentangle,gmb1dspinpol,gmb2d,gmb2dspinpol,gmb3dspinpol} as it simply decouples from the superconducting analysis and reduces the superconducting gap~\cite{gmb,gmb1dspinpol,gmb2d}. Magnetic fluctuations were neglected as these are small in spin-balanced systems.

A significant experimental consequence of communal pairing is the variation of chemical potential with scattering length, which may potentially be observed in the radius of trapped cold gases. In addition, other experimental techniques such as radio-frequency spectroscopy~\cite{trapfg1,trapfg2} that can directly probe the chemical potential. This reduction of the chemical potential compared to the BCS prediction may contribute to the persistent overestimation of the chemical potential by numerical methods compared to direct experimental measurements, such as those by the Jochim group~\cite{mumeasuregas}, with the magnitude of this mismatch being particularly well described by communal pairing theory near the unitarity limit of $\ln(\kF a)\approx1$, where interactions are sufficiently strong for effects to be visible beyond experimental uncertainty but still within the range of validity of the theory presented.

Another possible experimental signature is that the spatial structure of the superconducting gap should change with the scattering length, from isotropic in the weakly interacting limit to strongly confined in real space as interactions get stronger and the system approaches the BEC limit. This may be investigated in cold atomic gases, where control of the scattering length is well established \cite{trapfg1,trapfg2,uniformtrap}, for example using angle-resolved photoemission spectroscopy~\cite{gapmeasuregas}. The momentum-space structure of the gap could also be probed directly using Bogoliubov quasiparticle interference imaging~\cite{gapmeasuresolid}. In 2D, the analysis predicts that the superconducting gap has a width in momentum space that is inversely proportional to the scattering length, $q_\mathrm{2D}=\frac{2}{a}$ at weak interactions with $\kF a < 2$. In 3D, for weak interactions $\kF a_\mathrm{s} \to 0^-$, the width is predicted to follow $q_\mathrm{(3D)}\propto\kF \expe^{\pi/2\kF a_\mathrm{s}}$. However, we have demonstrated that low dimensionality promotes a higher $\Nq$ and so the experimental verification might be more straightforward in 2D systems. The additional pairing channels may also be visible through a range of retroreflected hole momenta in Andreev reflection experiments.

\section*{Acknowledgements}
Data used for this paper are available online~\cite{dspacefooconduit}. The authors acknowledge the financial support of the National University of Singapore and the Royal Society.

\bibliographystyle{unsrt}
\bibliography{bibo}

\end{document}